\documentclass[12pt]{iopart}
 \usepackage{iopams}
 \usepackage{epsfig}   
 \usepackage{graphics}
 \usepackage[T1]{fontenc}

\newcommand{\av}[1]{\langle{#1}\rangle}

\begin{document}

\title[Distance systematics and the CMB analysis]{The effect of inhomogeneities on the distance to the last scattering surface and the accuracy of the CMB analysis}

\author{Krzysztof Bolejko}
\address{Astrophysics Department,
University of Oxford,
1 Keble Road, Oxford OX1 3RH, UK}

\begin{abstract}
The standard analysis of the CMB data assumes that
the distance to the last scattering surface can be calculated using
the distance-redshift relation as
in the Friedmann model. However, 
in the inhomogeneous universe, even if $\av{\delta\rho} =0$,
the distance relation is not the same as in the unperturbed universe.
This can be of serious consequences as a change of distance affects the mapping 
of CMB temperature fluctuations into 
 the angular power spectrum $C_l$. In addition, if the change of distance is relatively uniform no new temperature fluctuations are generated.
It is therefore a different effect than the lensing or ISW effects
which introduce additional CMB anisotropies.
This paper shows that the accuracy of the CMB analysis can be impaired by the accuracy of calculation of the distance within the cosmological models.
Since this effect has not been fully explored before, to test how the inhomogeneities affect the distance-redshift relation, 
several methods are examined: the Dyer--Roeder relation, 
lensing approximation, and non-linear Swiss-Cheese model.
In all cases, the distance to the last scattering surface is
different than when homogeneity is assumed.
The difference can be as low as 1\% and as high as 80\%. 
Excluding extreme cases, the distance changes by about 20 -- 30\%.
Since the distance to the last scattering surface is set by the position
of the CMB peaks, in order to have a good fit, the distance needs
to be adjusted. 
After correcting the distance, 
the cosmological parameters change.
Therefore, a not properly estimated distance to the last scattering surface can be a major source
of systematics. This paper shows that 
if inhomogeneities are taken into account when calculating the distance
then models with positive spatial curvature and with $\Omega_\Lambda \sim 0.8-0.9$ are preferred. The $\Lambda$CDM model (i.e. a flat Friedmann solution with the cosmological constant), in most cases, is at odds with the current data.
\end{abstract}

\section{Introduction}

It has been said that we entered the era of precision cosmology. This is mainly due to observations
of the cosmic microwave background radiation (CMB). 
As the CMB power spectrum  is very sensitive to cosmological parameters it
provides very tight constraints. Currently the errors are at the level of a few percent \cite{WMAP7}. 
The CMB power spectrum is shaped by several processes that can be divided into two groups.
The first group involves processes that occurred
before and during the generation of the CMB.
This part is well understood as the Universe at that times was very close to homogeneous 
and contribution from the spatial curvature and dark energy was negligible.
The constraints on cosmological parameters coming just from this group of processes
ware recently presented and discussed in \cite{VoRD2010}.
The second group of processes takes into account what happens with photons between the last scattering surface 
and the observer. This includes the distance to the last scattering surface,
the ISW effect, the impact of reionization on the CMB, the Sunyaev-Zel'dovich
and the lensing effects.

The analysis of this second group of effects is a subject to systematics, as the geometry of the late time Universe does not have
to be close to homogeneous and isotropic Robertson-Walker geometry
(these processes are most often studied within the framework of linear perturbations around the homogeneous Friedmann models).
On small scales (say $\ell > 60$, i.e. $\theta<3^\circ$), expect for the distance, these processes should not significantly modify the overall shape of the CMB power spectrum. The distance to the last scattering surface affects the mapping 
of CMB temperature fluctuations into the angular power spectrum $C_l$.
This effect can be modelled by introducing the shift parameter ${\cal R}$, which is proportional to the distance to the last scattering surface \cite{BoET1997,EfBo1999}.
Therefore, there is a number of papers where the constraints from the CMB
are just limited to the constraints on the shift parameter.
This is usually done when testing different models of dark energy \cite{MMOT2003,KuSz2008},
alternative cosmological models \cite{LaMM2006,RyFG2007}, 
or Gpc-scale inhomogeneous models \cite{AAG06,AlAm2007,BoWy2009,ClFZ2009,ABNV2009,ZaMS2008,ClRe2010,YoNS210,MZS10,BNV10}.
As discussed in \cite{WaMu2007,ElMu2007} this type of analysis can be improved but taking into account both the scale of the sound horizon at the last scattering and ${\cal R}$.

This paper studies the effect of the distance but does not consider
 any alternative cosmologies.
Here we just focus on the effect of small-scale inhomogeneities on the distance to the last scattering.
The presence of inhomogeneities is know to affects the distance-redshift relation 
\cite{Sach1961,KrSa1966}. Some studies claim that the effect is large
\cite{KaVB1995,MKMR2007,MaKM2008,KaMa2009}, others that it is small 
\cite{BrTT2007,BrTT2008,VaFW2008,WV09,Szyb2010}.
Even if the change of the distance just at the level of a few percent 
\cite{ClZu2009,ClFe2009,kbAA,kbMNRAS} this analysis is still important 
as a change of the distance by a few percent leads  
to a similar change of the inferred values of cosmological parameters.

One may think that the effect of inhomogeneities is taken into account in the standard
analysis of the CMB via the ISW or lensing effects. 
However, in both cases (in the standard approach) this is done by using the matter power spectrum. Thus, this analysis is insensitive to the change of the mean, i.e. the uniform change of the distance to the last scattering surface.
For example the lensing analysis deals with the change of the distance but
is only sensitive to the change of the variance, not the mean (cf.  \cite{Lcmb1,Lcmb2}).
If the change of distance is relatively uniform (i.e. with a negligible variance) then no additional temperature anisotropies are generated.
Since the change of distance affects the mapping of the physical position of the peaks to the peaks in the angular power spectrum $C_l$ this effect needs to be considered also at the level of the mean.

Therefore, if inhomogeneities are not taken into account
when calculating the distance, then they may become a major source of systematics.
With the increasing precision of CMB experiments, in particular Planck, 
a proper understanding of systematics is important. 
Without taking into account all systematics the precision cosmology will not be an accurate cosmology.
Therefore, this paper aims to study 
how the presence of inhomogeneities affects the distance-redshift relation, in particular the distance to the last
scattering surface.
Several models are considered and it is shown that the change of distance due to inhomogeneities is not negligible.

The structure of this paper is as follows: Sec. \ref{distance} discusses
different methods for the distance calculation, Sec. \ref{dissec} applies these methods 
to calculate the distance to the last scattering surface, Sec. \ref{labdc} explores the implication of differences in the distance, and Sec. \ref{conk} concludes the results.

\section{Distance and distance corrections}\label{distance}

As shown by Sachs \cite{Sach1961}, the equation for the angular diameter distance
$D_A$
is
\begin{equation}
\frac{{\rm d^2} D_A}{{\rm d} s^2} = - ( |\sigma|^2 + \frac{1}{2} R_{\alpha
\beta} k^{\alpha} k^{\beta}) D_A, \label{dsr}
\end{equation}
where $s$ is a parameter along the geodesic, $k^\alpha$ is a null vector that is tangent to the geodesic,
$\sigma$ is the shear of the light bundle,
$R_{\alpha \beta}$ is the Ricci tensor and $R_{\alpha \beta} k^\alpha k^\beta =
\kappa T_{\alpha \beta} k^\alpha k^\beta$. In the
comoving and synchronous coordinates, for pressure-less matter,  $R_{\alpha \beta} k^\alpha k^\beta =
\kappa \rho k^0 k^0$. 
As seen, the distance depends on density
fluctuations, and secondly, even if
the perturbations vanish after averaging (i.e. $\av{\rho} = \av{\rho_0 + \delta
\rho} = \rho_0$, where $\rho_0$ is the background density) they do modify the
distance relation and the final result deviates from the
homogeneous solution.
Equation (\ref{dsr}) is the differential equation
so it is sensitive to any change of initial conditions.
Therefore, it is sensitive to 
$\delta \rho D_A$ as this term acts as a potential (for a detailed discussion see \cite{kbAA}).
As the parameter $s$ is not directly measured (unless the geodesic parametrization
is chosen in a particular way), it is convenient to express it in terms of a redshift. 
The redshift is given by the following relation \cite{Elli1971}

\begin{equation}
1+z = \frac{(u_\alpha k^\alpha)_e}{(u_\alpha k^\alpha)_o},
\end{equation}
where $u^\alpha$ is the 4-velocity, and subscripts $e$ and $o$ refer to emission and observation instants respectively.
In the comoving and synchronous coordinates $u^\alpha = \delta^\alpha_0$ and  
\[1+z = \frac{k^0_e}{k^0_o}, \]
where $k^0 = {\rm d} t/{\rm d s}$. Thus, for a given model and using the above relation the parameter $s$ can be 
replaced with $z$.

\subsection{Homogeneous universe}

When homogeneity is assumed, i.e. $\sigma = 0 = \delta \rho$, {  and $k^0 = 1/a$ (where $a(t)$ is the scale factor)
then ${\rm d} z / {\rm d} s =  - (1+z)^2 H_0 \sqrt{ \Omega_{m} (1+z)^3+\Omega_{k}(1+z)^2 + \Omega_{\Lambda} }$ and 
(\ref{dsr}) reduces to}

\begin{equation}
 D_{A}(z)=\frac{1}{H_0 (1+z) \sqrt{-\Omega_k}}\sin{\left(
\int_0^z{\mathrm{d}z'\frac{\sqrt{-\Omega_k}}{
\sqrt{ \Omega_{m} (1+z)^3+\Omega_{k}(1+z)^2 + \Omega_{\Lambda} }
}}\right)},
\label{dlcdm}
\end{equation}
where $H_0$ is the Hubble constant, 
$\Omega_k = - k/( H_0^2 a_0^2)$,
$\Omega_m = (8\pi G \rho_0)/ (3H_0^2)$, and 
$\Omega_{\Lambda} = \Lambda / (3H_0^2)$.

\subsection{Dyer--Roeder approximation}

The Dyer--Roeder approach also assumes homogeneity 
but takes into account that light propagates through vacuum. Therefore,
 $\Omega_m$ which photons `feel' is different than true $\Omega_m$. 
This is modelled by a constant parameter $\alpha$
(of value between 0 and 1) that multiplies  $\Omega_m$.
In this case (\ref{dsr}) reduces to \cite{DyRo1972,DyRo1973,Matt2010}

\begin{equation}
\frac{{\rm d}^2 D_A}{{\rm d} z^2} + 
\left(\frac{1}{H}\frac{{\rm d} H}{ {\rm d} z} + \frac{2}{1+z} \right) \frac{{\rm d} D_A}{ {\rm d} z}
 + \frac{3}{2} \frac{H_0^2}{H^2} \alpha  \Omega_m  (1+z)  D_A=0,
\label{ddre}
\end{equation}
where
$ H(z)=H_0 \sqrt{\Omega_{m} (1+z)^3+\Omega_{k}(1+z)^2 + \Omega_{\Lambda}}.$
In order to include the evolution of density fluctuations along the line of sight,
Ref. \cite{kbMNRAS} suggests the following modification of the Dyer--Roeder equation

\begin{equation}\label{mddre}
\frac{{\rm d}^2 D_A}{{\rm d} z^2} + 
\left(\frac{1}{H}\frac{{\rm d} H}{ {\rm d} z} + \frac{2}{1+z} \right) \frac{{\rm d} D_A}{ {\rm d} z} +  \frac{3}{2} \frac{H_0^2}{H^2} \Omega_m (1+z)
\left(1 +  \frac{\av{\delta}_{1D}}{(1+z)^{5/4}} \right)
 D_A=0,
\end{equation}
where $\av{\delta}_{1D}$ is the mean of the present-day density fluctuations along the line of sight. 

Initial conditions needed to solve (\ref{ddre}) and (\ref{mddre}) are:
$D_A(0) = 0$ and  ${{\rm d} D_A}/{ {\rm d} z} = 1/H_0$.

\subsection{Linear perturbations and the lensing approximation}

Within the linear perturbative scheme the distance is
\begin{equation}
 D_{A}(z)= \bar{D}_{A} ( 1 + \Delta_D),
\label{dlen}
\end{equation}
where $\bar{D}_{A}$ the distance in the homogeneous universe
and is given by (\ref{dlcdm}). 
The most general form for $\Delta_D$ is presented in \cite{PyBi2004,BDG2006,HuGr2006,EnMR2009}.
Neglecting the contribution from the observer's and source's motion, and taking the leading term, $\Delta_D$ can be expressed as

\begin{equation}\label{dDBa}
\Delta_D =
- \int\limits_0^{\chi_e} {\rm d} \chi
\frac{ \chi_e - \chi}{\chi_e} \chi \nabla^2 \phi(\chi),
\end{equation}
where $\chi$ is the comoving coordinate $d \chi = dz/H(z)$,
$\phi$ is the gravitational potential that
relates to the density perturbations $\delta \rho$ via the  Poisson equation
$ \nabla^2 \phi = \frac{4 \pi G}{c^2} a^2 \delta \rho$.
The above expression is equivalent to the 
convergence in the lensing approximation and is known as the Born approximation.
As seen, propagation through voids ($\delta \rho<0$) increases the distance
while propagation through overdense regions ($\delta \rho>0$) decreases the distance.

In order to solve (\ref{dDBa}) one needs to know exactly what the density fluctuations along the line of sight are. 
However, within the weak lensing analysis one does not solve (\ref{dDBa}) directly.
Instead the variance is calculated, thus instead of $\delta \rho$ 
one has $(\delta \rho)^2$ which can be expressed in terms of the matter power spectrum.
In this paper we are interested in the overall (uniform) change of the distance.
Thus, we will solve (\ref{dDBa}) directly, this requires 
knowledge of density fluctuations along the line of sight. The algorithm
is described below, but see also \cite{KaMa2009,KaMa2010} for an alternative approach.

\section{Distance to the last scattering surface}\label{dissec}

\begin{figure}
\begin{center}
\includegraphics[scale=0.7]{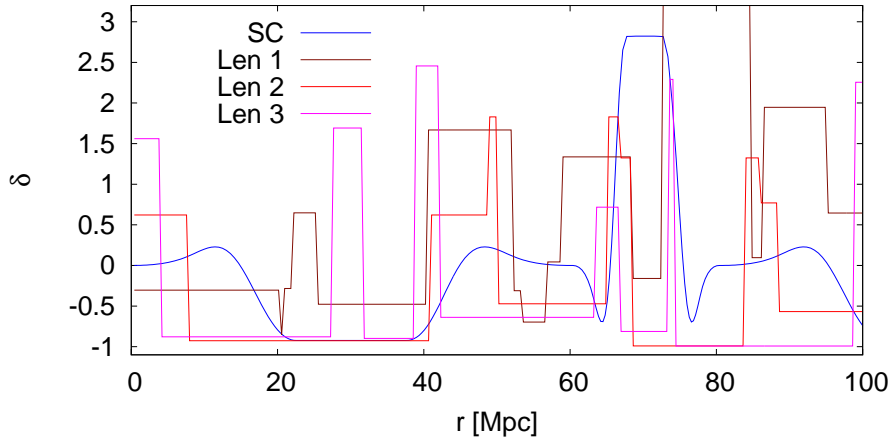}
\caption{Density fluctuations along the line of sight within the lensing approximation and
the Swiss Cheese model.} \label{fig1}
\end{center}
\end{figure}

To test what is the effect of inhomogeneities on the distance to the last scattering
surface and the accuracy of the CMB analysis, let us consider 7 different
ways of calculating the distance to the last scattering surface

\begin{enumerate}
\item The $\Lambda$CDM model:

Distance to the last scattering surface is calculated from (\ref{dlcdm}).
\item Dyer--Roeder approximation (model DR):

Based on the cosmological observations, Yu et al. \cite{YLWZW2010} suggest that $\alpha=0.93$. Therefore, the distance to the last scattering surface
is calculated using (\ref{ddre}) with $\alpha=0.93$.
\item Modified Dyer--Roeder relation (model mDR):

The modified version of the Dyer--Roeder equation (\ref{mddre}) is solved
with $\av{\delta}_{1D} = -0.3$.
\item Lensing approximation  (model Len 1):

In order to solve (\ref{dDBa}), one needs to know $\nabla^2 \phi$ along the line of sight. Using the Poisson equation the gravitational potential is related to density fluctuations. 
The density fluctuations along the line of slight are generated from the log-normal PDF -- the algorithm is discussed in details in \ref{Alen1}.
\item Lensing approximation (model Len 2):

In model Len 1 the mean of density fluctuations
along the line of sight is negligible. However, as pointed out in \cite{kbMNRAS} the mean of density fluctuations along the line of sight does not have to vanish, as $\av{\delta}_{3D}=0$ (3D -- volume average) does not necessarily imply $\av{\delta}_{1D}=0$ (1D -- along the line of sight). In model Len 2,  $\av{\delta}_{1D}\ne0$ 
although $\av{\delta}_{3D}=0$. The algorithm is discussed in details in  \ref{Alen2}.
\item Lensing approximation  (model Len 3): 

Model Len 3 uses a more flexible (than Len 1 \&  2) method of generating density fluctuations along the line of sight.
The algorithm is presented in details in \ref{Alen3}.

\item Fully non-linear model (model SC):

The above approaches were based either  on the assumption of homogeneity (the $\Lambda$CDM and Dyer--Roeder relation) or on the linear approximation (models Len 1--3).
In fact, the lensing approximation is not even fully linear, as it takes into account
just the leading linear term. Also, when generating fluctuations
along the line of sight, the matching conditions were not taken into account,
i.e. the structures were not properly matched. In addition, the light rays
were not properly match as well, i.e. when light ray exits one structure
and enters another one, the null geodesics need to be properly matched.
Therefore, a fully non-linear model is considered here, where
the distance is calculated directly from (\ref{dsr}) 
and all matching conditions are handled properly.
The model is based on the Swiss-Cheese model. Each inhomogeneous patch is of  spherical symmetry and is  described using the
Lema\^itre-Tolman (LT) model \cite{Lema1933,Tolm1934}. Thus, density within each
inhomogeneous patch changes continuously, unlike in the lensing approximation, where within each generated structures, matter density is constant. 
Also the evolution of matter is calculated using the exact method, not the linear 
approximation. 
The properties of the model are chosen so that the mean of density
fluctuations along the line of sight is zero, i.e. $\av{\delta}_{1D} \approx 0$.
Thus, this model allows us to test how large non-linear contributions
are, in contrast to $\av{\delta}_{1D} \ne 0$ contributions.  The detailed 
description of this model is presented in \ref{Asc}.
\end{enumerate}

\begin{figure}
\begin{center}
\includegraphics[scale=0.7]{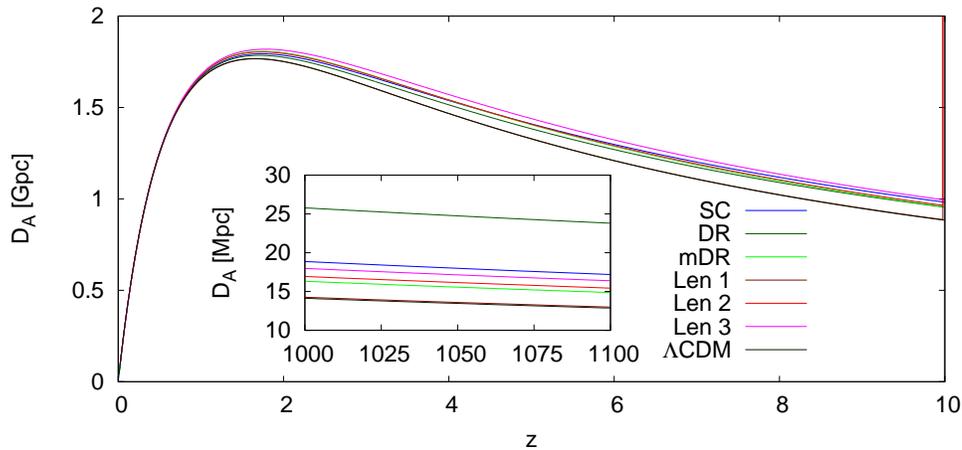}
\caption{The angular diameter distance within 7 considered models. The inset
presents $D_A$ in the vicinity of the last scattering surface.} \label{fig2}
\end{center}
\end{figure}

In this section it will be assumed that the large-scale universe is exactly the $\Lambda$CDM,
so $H_0 = 71.4$ km s$^{-1}$ Mpc$^{-1}$, $\Omega_m = 0.262$,
$\Omega_\Lambda = 0.738$.
 A sample of 
density fluctuations along the line of sight is  presented in Figure \ref{fig1}. 
The results, in the form of $D_A(z)$, are presented in Figure \ref{fig2}.
The distance to the last scatting surface is  presented in the inset,
and in the form of the shift parameter ${\cal R}$ in the 2nd column of Table \ref{tab1}.
As seen, the distance to the last scattering surface depends on the choice of the method.
Even within the most optimistic case, model Len 1, the change is of order of $1\%$.
However, the non-linear contributions (even if $\av{\delta}_{1D} \approx 0$)
gives $\sim 30\%$ change of the distance. 
When $\av{\delta}_{1D} \ne 0$, even  within the lensing approximation
the change is of order of $20-30\%$.

Another important fact is that the variance within each method is small.
Models Len 1--3 allow for random (within the limits of the method)
density fluctuations along the line of sight. Based on 100,000 runs
(each with a different distribution of density fluctuations along the line of sight)
the results are 
Len 1: $\Delta_D =  -0.0071 \pm 0.0198$, Len 2: $\Delta_D =  0.2033 \pm 0.0064$, Len 3: $\Delta_D =  0.2729 \pm 0.0071$. This shows that the change of distance is relatively uniform. Below, when referring to models 
Len 1-3 it will be referred to the mean of $\Delta_D$ \footnote{It should be noted that the distance corrections depends on 
cosmological parameters of the background model, for example 
$\Delta_D$ for model Len 1 changes to $-0.0064$ and $-0.0069$
when parameters are as in Table \ref{tab1} and Table \ref{tab2} respectively.}.

In the next section we will study the implications
of the change of the distance to the last scattering surface and the
CMB constraints when this change is taken into account.

\section{CMB analysis with the adjusted distance-redshift relation}\label{labdc}

\subsection{Fitting the distance and the accuracy of CMB analysis}

In this section we assume that the physical densities of baryonic and cold dark matter, 
and primordial power spectrum are the same as in the standard cosmological model, and 
only implications of the change of the distance to the last scattering surface are going to be examined.

\begin{figure}
\begin{center}
\includegraphics[scale=0.7]{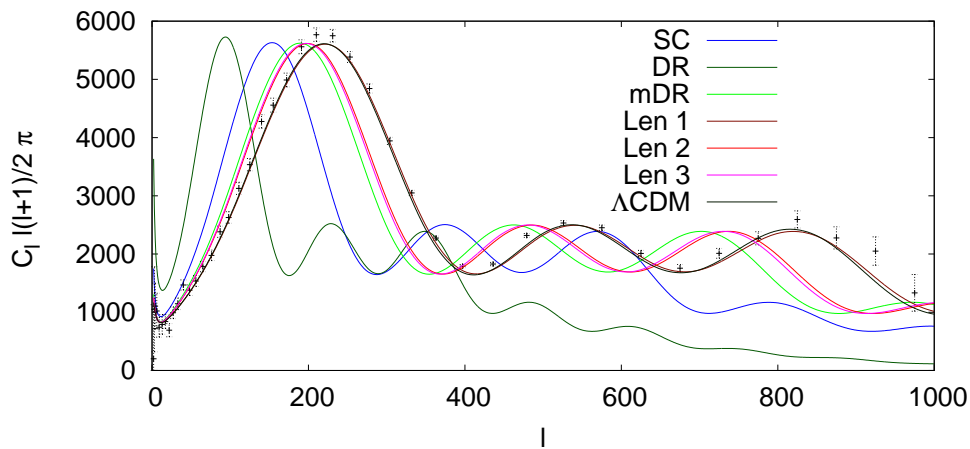}
\caption{The CMB power spectrum calculated with CAMB 
\cite{CAMB} for homogeneous cosmological models with parameters as in Table \ref{tab1}.} \label{fn1}
\end{center}
\end{figure}

\begin{figure}
\begin{center}
\includegraphics[scale=0.7]{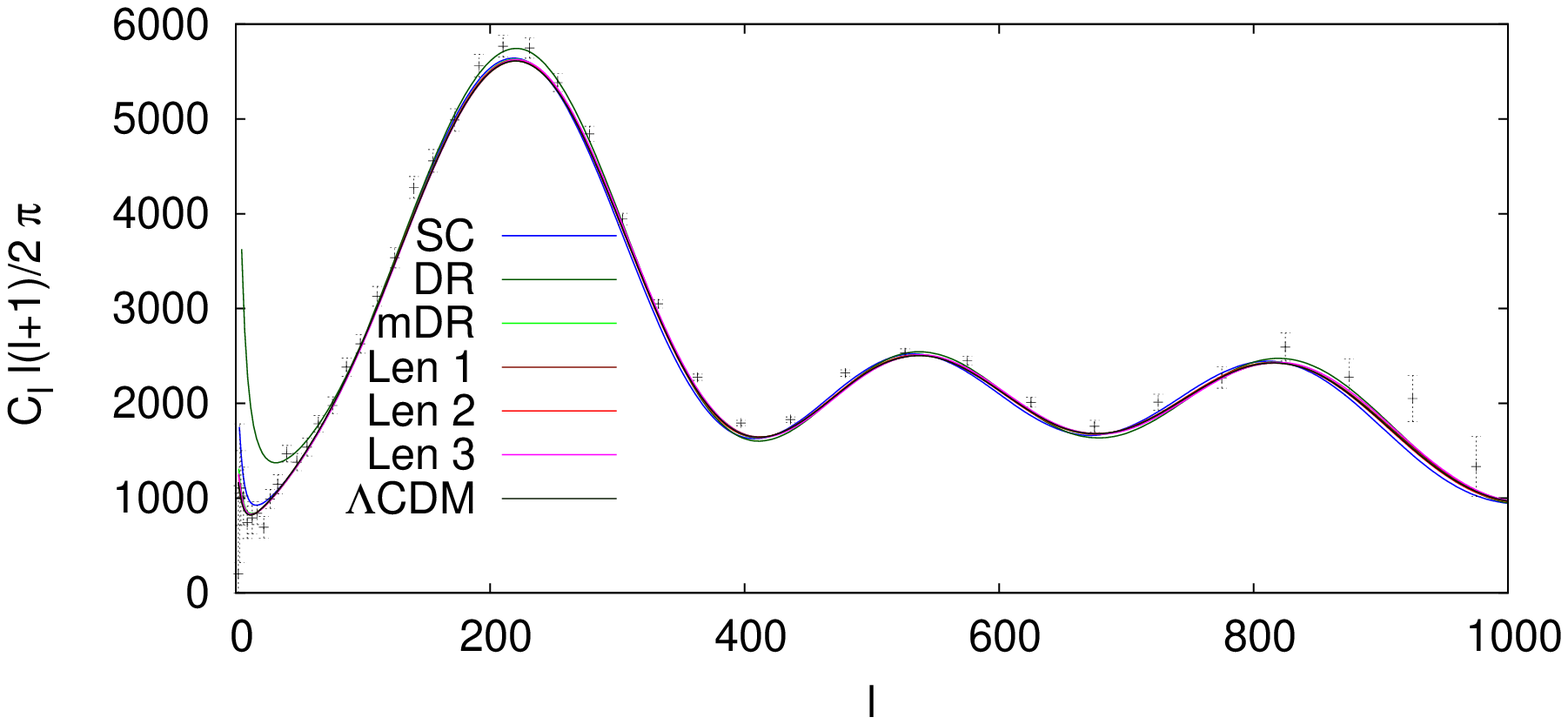}
\caption{The CMB power spectrum calculated with CAMB 
\cite{CAMB} and corrected for the distance (i.e. 
with the scaling (\ref{cmbscaling}) employed).
 The inset presents the power spectrum for low-$\ell$ with errors
enlarged to include the cosmic variance.} \label{fig3}
\end{center}
\end{figure}

The change of the distance-redshift relation leads to a different
mapping of the CMB temperature fluctuations into 
the angular power spectrum $C_l$.
Thus, if the distance-redshift relation changes,
then the standard analysis of the CMB needs to be adjusted.
The way to do this is following: first the CMB power spectrum is calculated
using standard codes (for example CAMB \cite{CAMB}) then the output 
requires following scaling

\begin{equation}
\ell\rightarrow \ell (1+\Delta_D) \quad {\rm ~and~} \quad C_\ell \rightarrow C_\ell (1+\Delta_D)^{-2}.
\label{cmbscaling}
\end{equation}
Thus, the $\ell$ axis changes while the
amplitude $\ell (\ell +1 ) C_\ell$ remains relatively unchanged
(for a detailed discussion see \cite{ClFZ2009,VoRD2010,ClRe2010}).
To present how it works
let us consider the following example: let us change 
the value of $\Lambda$ in such a way so that the distance
to the last scattering instant, for studied models, is
the same as in the standard cosmological model.
$\Omega_\Lambda$ needed to fit the distance, for studied models,
is presented in the 3rd column of Table \ref{tab1}.
Apart from model Len 1, $\Omega_\Lambda$ is much larger than
when homogeneity is assumed. 
If the power spectrum was calculated for a homogeneous Friedmann model
(i.e. with the distance-redshift relation the same as in the Friedmann model)
the result would look like the one presented in Figure \ref{fn1}.
However, after applying scaling (\ref{cmbscaling})
the angular power spectrum is as in Figure \ref{fig3}. 
As seen the CMB power spectrum is very similar except for low-$\ell$ where the ISW effect
is different. The low-$\ell$ power spectrum is presented in the inset, where the error-bars
are enlarged to include the cosmic variance $\Delta C_\ell / C_\ell = \pm (2/(2 \ell + 1))^{1/2}$.

\begin{figure}
\begin{center}
\includegraphics[scale=0.7]{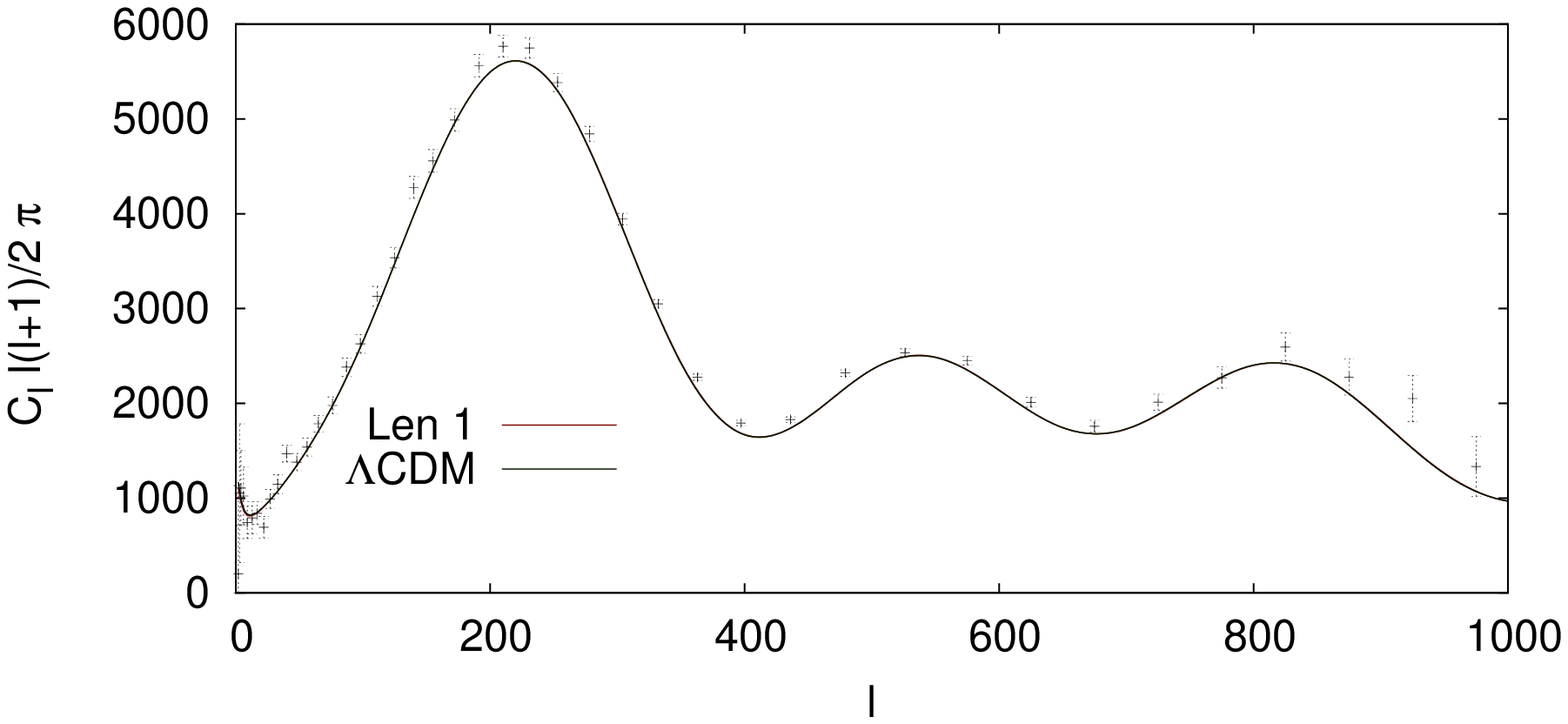}
\caption{The CMB power spectrum calculated with CAMB 
\cite{CAMB} and corrected for the distance (i.e. 
with the scaling (\ref{cmbscaling}) employed). The inset presents the power spectrum for low-$\ell$ with errors
enlarged to include the cosmic variance.} \label{fig4}
\end{center}
\end{figure}

Another example: let us  correct the distance by
changing the cosmological parameters but keeping
the spatial curvature zero, i.e. 
$\Omega_m + \Omega_\Lambda = 1$.
Thus, the change of $\Omega_m$ automatically implies the change of $\Omega_\Lambda$.
Also to keep the shape of the CMB power spectrum unchanged,
$\Omega_m h^2 = (\Omega_b + \Omega_c) h^2$ must not change, so the 
change of $\Omega_m$ automatically implies the change of $H_0$.
The change of cosmological parameters needed to fit the distance
is presented in Table \ref{tab2}.
As seen all models except for model Len 1 can be ruled out because of too
high amplitude of $H_0$. Even within model Len 1, the change of 
parameters is large compared to the current precision -- 
for example in model Len 1, $\Omega_m$ is around $0.28$
not $0.262$ so the difference is around 7\%, 
which is twice as large as the error estimated by the WMAP team \cite{WMAP7}.
For completeness the CMB power spectrum for Len 1 and the $\Lambda$CDM model
are presented in Figure \ref{fig4}.

The above examples show why it is important to have an accurate method
for calculating the distance.
As the universe is inhomogeneous, the distance-redshift relation 
is different than the distance-redshift relation in the Friedmann model.
As shown, even a small change of the distance 
can have a large impact on the accuracy of the CMB analysis.
As seen from Table \ref{tab1} a change of the distance by $1\%$ leads to the 0.5\% 
change of the cosmological constant. Although this change is small, in the 
era of precision cosmology this may be important.
Especially that as shown in Table \ref{tab2} if one puts by hand the assumption
of spatial flatness, then systematics are much larger.

\begin{table}
\begin{center}
\caption{The effect of inhomogeneities on the distance to the last scattering surface -- 2nd column
(${\cal R} = D_A (1+z) \sqrt{\Omega_m} H_0/c$);
the accuracy of estimation of $\Lambda$ -- 3rd and 4th column.
In all models $H_0 = 71.4$ km s$^{-1}$ Mpc$^{-1}$, $\Omega_c h^2 = 0.1107$, $\Omega_b h^2 = 0.0227$.}
\begin{tabular}{lccc}
 \hline
 Model & the shift parameter  ${\cal R}$  & $\Omega_\Lambda$  needed to & $
\delta_\Lambda = | \Delta \Omega_\Lambda| / \Omega_\Lambda$    \\
 & (if the background model   & fit the CMB & \\
 & is the $\Lambda$CDM model) &              &        \\
 \hline
$\Lambda$CDM & 1.73  & 0.738 & 0.0  \\
       Len 1 & 1.71  & 0.734 & 0.005   \\
       Len 2 & 2.08  & 0.797 & 0.08  \\
       Len 3 & 2.19  & 0.802 & 0.09  \\
          SC & 2.31  &  0.919 & 0.24  \\
          DR & 3.19  & 1.071 & 0.45  \\
         mDR & 2.04 & 0.823  & 0.12  \\
\hline
\label{tab1}
\end{tabular}
\end{center}
\end{table}

\begin{table}
\begin{center}
\caption{The effect of inhomogeneities 
on the accuracy of inferred cosmological parameters.
In all models $\Omega_c h^2 = 0.1107$, $\Omega_b h^2 = 0.0227$, and $\Omega_m + \Omega_\Lambda = 1$.}
\begin{tabular}{lcccccc}
 \hline
 Model & $\Omega_m$ & $\delta_m = | \Delta \Omega_m| / \Omega_m$ & $\Omega_\Lambda$ & $\delta_\Lambda = | \Delta \Omega_\Lambda| / \Omega_\Lambda$ & $h$ & $\delta_h = | \Delta h| / h$ \\
$\Lambda$CDM &  0.262 & 0 & 0.738 & 0 & 0.714 & 0 \\
Len 1 &  0.281 & 0.07 & 0.719 & 0.03 & 0.690 & 0.03 \\
Len 2 &  0.050 & 0.81 & 0.95 & 0.29 & 1.64 & 1.30 \\
Len 3 &  0.034 & 0.87 & 0.966 & 0.31 & 1.982 & 1.78 \\
SC &     0.024 & 0.91 & 0.976 & 0.32 & 2.344 & 2.28 \\
DR &     0.006 & 0.98 & 0.994 & 0.35 & 4.568 & 5.40 \\
mDR &    0.095 & 0.64 & 0.905 & 0.23 & 1.19 & 0.66 \\
\hline
\label{tab2}
\end{tabular}
\end{center}
\end{table}

\subsection{Spatial curvature of the Universe}\label{curv}

Let us now focus only on models Len 1, Len 2, and SC.
We exclude the Dyer--Roeder relation as this is only a rough approximation (as the shear accumulates along geodesics this approximation
is expected to be inaccurate for large distances). 
Also as model Len 3 is rather of arbitrary matter fluctuations
(it was considered only to check how arbitrariness of matter fluctuations changes 
the results) it is excluded from the further and more detailed analysis.

In this section we will only focus on Len 1, Len 2, and SC models.
Len 1 model is a conservative approach  -- it is based on the linear approach
with random density fluctuations along the line of sight. Len 2 model 
is also based on linear approximation, but the mean of density fluctuations
along the line of sight is non-zero (as discussed in \cite{kbMNRAS} if structures
in the universe are organized, then density fluctuations are no longer purely random, hence
even though the 3D average vanish, the average of density fluctuations over the 
the line of sight is not zero).
SC model is a fully non-linear model with vanishing density fluctuations along the line of sight.

\begin{figure}
\begin{center}
\includegraphics[scale=0.85]{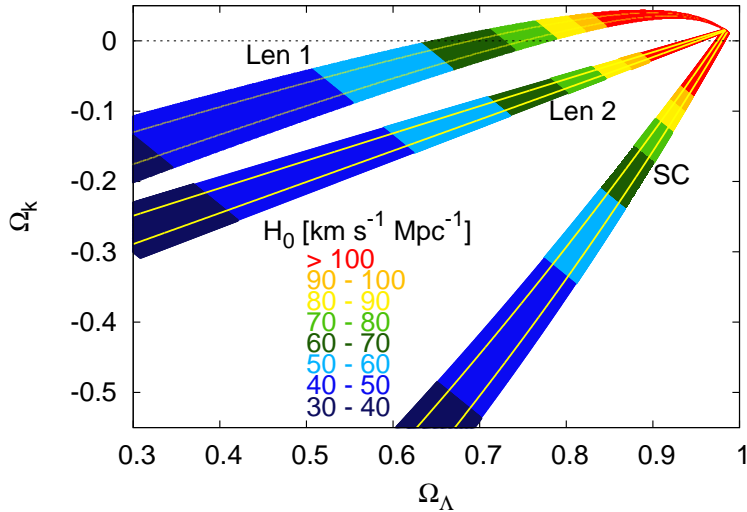}
\caption{The CMB constraints after taking into account the effect of inhomogeneities
on the distance to the surface of last scattering, for Len 1, Len 2, and SC models. The color regions 
are the 95\% confidence interval, the yellow lines indicate the 68\% confidence interval.} \label{crh}
\end{center}
\end{figure}

The results of the CMB analysis within these 3 models is presented in Fig. \ref{crh}.
Now we allow $\Omega_m$ and $\Omega_\Lambda$ to vary freely (however, we keep
$\Omega_b h^2$ and $\Omega_c h^2$ unchanged, so the change of $\Omega_m$
implies the change of $H_0$).
As seen the constraints obtained within Len 1 model are very similar 
to the constraints obtained within the standard cosmological model.
However, model Len 1  does not include non-linear corrections. 
If these are taken into account or if only the mean of density
fluctuations along the line of sight does not vanish then
spatial curvature cannot be zero. As seen 
the presence of small-scale non-linear structures changes the distance-redshift relation
and now the CMB data combined with local value of $H_0$ no longer favors spatially flat models.
This shows that the flat Friedmann solution is not preferred by the data, and when inhomogeneities are included into the distance-redshift relation
then $k>0$ models with larger amount of dark energy are favored.

Let us discuss for a moment the possibility that 
in fact $\Omega_\Lambda \approx 0.8$.
What are the implications? In this case,
the age of the universe increases to approximately $14$ Gyr.
This change is acceptable. An increased amount of dark energy should be
detectable by other cosmological observations. The supernova data
alone are not very precise, but still 
homogeneous Friedmann models that fit the data best are 
the elliptic, not flat, models with $\Omega_\Lambda \approx 0.9$ \cite{ALRA2010}. 
Also, the BAO analysis is in the agreement with larger amount dark energy.
However, it should be noted that one of the elements of the BAO 
analysis is a conversion from the redshift to distance. This conversion is obtained 
using the Friedmann models. Therefore, if the
distance-redshift relation is changed then the result of the BAO analysis can also
be different. This should be kept in mind when analyzing the BAO data, especially that
as shown in \cite{kbAA} ignoring inhomogeneities also leads
to additional systematics.

\section{Conclusions}\label{conk}

This paper investigated how the presence of inhomogeneities 
(small-scale inhomogeneities observed in the Universe)
affects the distance to the surface of last scattering,
and how via the change of the distance the inhomogeneities influence the CMB data.

In the standard approach, the distance is calculated within a framework of homogeneous
Friedmann models. There are several reasons for using this approach.
First of all,
there is an argument used by Weinberg \cite{Wein1976} who pointed out
that although for a single case the distance is modified
by the inhomogeneities, but  
due to photon conservation,
when averaged over large enough angular scales 
the overall effect is zero (however see \cite{ElBD1998}
for an argument why Weinberg's reasoning should not apply).
Second of all, as seen from (\ref{dDBa})
if the amount of voids is the same as amount of overdensities
then $\Delta_D$ should be zero and the distance should be exactly the same
as in the Friedmann model. 
As density fluctuations should vanish after averaging over sufficiently large scales
it seems reasonable to expect that the Friedmann distance-redshift relation  
correctly describes the reality.

However, there is a difference between
vanishing 3D average and vanishing 
of the average of the density fluctuations over the line of sight.
If density fluctuations in the Universe are purely random
then vanishing 3D average implies vanishing average over the line of sight.
If density fluctuations are not purely random, i.e. if there is some
degree of organization, then vanishing  3D average does not necessarily
imply vanishing density fluctuations over the line of sight
(for a discussion see \cite{kbMNRAS}).
As the large-scale matter distribution in the Universe has a form of the cosmic web,
there is some kind of organization, and thus the result of (\ref{dDBa})
does not have to be zero, even if 3D average of density fluctuations 
vanishes after averaging over sufficiently large scales. 
This is the case of model Len 2.
Also, relation (\ref{dDBa}) does not apply when structures are non-linear.
The calculations within model SC were carried out within the non-linear regime.
In addition matter distribution was chosen so that the density fluctuations
along the line of sight vanish after averaging (i.e. $\av{\delta}_{1D} = 0$).
In this case, again, the distance was not the same as 
within the Friedmann model.

Even if the change of the distance is small, like in Len 1 model,
still the effect is important, as we have already entered the era of precision cosmology --
as seen from Table \ref{tab1} and Table \ref{tab2} the change 
of the inferred cosmological parameters, even for Len 1 model, can be of order of a percent or more.
If the change of the distance is larger, for example as in models Len 2 and SC, then 
the inferred parameters are significantly different.
In this case spatially flat models are ruled out and the data
favors positively curved models with with $\Omega_\Lambda \approx 0.8-0.9$.

As discussed in Sec. (\ref{dissec}),
within a chosen method, the 
distance to the last scattering surface is relatively uniform 
with only small variance.
The presented, in this paper, results show 
that the change of the mean is important, and the proper
handling of this effect is essential, otherwise the systematics
may be larger than the precision of measurements.
The only problem, however, is to have a good method for the calculation of the distance.
All models presented in this paper have their limitations, and should
rather be treated as toy models, i.e. as examples to show the significance of the problem.
The results show that the accuracy of the CMB analysis strongly depends on the
 accuracy of the  calculation of the distance corrections.
Peebles described the importance of accuracy and precision in cosmology with the following example
{\it a digital scale may read out the weight of an object to many significant figures, in a precise measurement. But if the scale is not well calibrated the measurement may not be very accurate} \cite{Peeb2010}.
In our case, the `calibration' requires 
an inhomogeneous framework that will allow us to estimate the distance
with inhomogeneities taken into account. As the non-linear corrections are important
this cannot be done within the lensing approximation (\ref{dDBa})
but requires a fully inhomogeneous framework.
Only then we will be able to accurately predict the distance and say:
`the distance is larger by $29.72\%$' or `$36.14\%$ than when homogeneity is assumed'.
With the increasing precision of CMB experiments, in particular Planck, 
Atacama Cosmology Telescope, and South Pole Telescope this is important. Otherwise,
what would be the meaning of a measurement with 
very high precision ($\sim 1\%$ statistical errors) 
 if the accuracy is low ($\sim 30\%$ systematics)?

\ack

I would like to thank Richard Bond, 
Chris Clarkson, Timothy Clifton, Ruth Durrer, Pedro Ferreira, Valerio Marra,
Teppo Mattsson, Syksy R\"as\"anen, Marco Regis, David Wiltshire for useful discussions and suggestions.
This research of was supported by 
the Marie Curie Fellowship (PIEF-GA-2009-252950).

\appendix

\section{Construction of model Len 1}\label{Alen1}

In order to solve (\ref{dlen}) one needs to know $\nabla^2 \phi$ along the line
of sight. Using the Poisson equation the gravitational potential is related to density fluctuations. Therefore, a method of generating density fluctuations is needed.
In the non-linear regime,
the density fluctuations 
 are not Gaussian (there is no symmetry as 
$-1 \leq \delta < \infty$). However, 
in the non-linear regime the distribution of density fluctuations can  be approximated  with 
the one-point log-normal PDF \cite{KaTS2001,LaSu2004}

\begin{equation}\label{nlPDF}
P(\delta) = \frac{1}{\sqrt{2 \pi \sigma_{nl}^2} } 
\exp \left[ - \frac{ (\ln(1+\delta) + \sigma_{nl}^2/2)^2 }{2 \sigma_{nl}^2} \right] \frac{1}{1+\delta},
\end{equation}

where
\begin{equation}\label{nlVar}
\sigma_{nl}^2 = \ln [ 1 + \sigma_R^2], \quad {\rm ~and~} \quad
\sigma_R^2 = \frac{1}{2\pi^2} \int\limits_0^\infty {\rm d} k {\cal P}(k) W^2(kR) k^2,
\end{equation}
and ${\cal P}(k)$ is the matter power spectrum.
Therefore, the algorithm for $\delta$ along the line of sight is as follows:

\begin{enumerate}
\item
The radius of a structure $R$ is randomly generated from a uniform distribution,
from 0 to 10 Mpc.
\item
Then $\sigma_R$ is calculated using  (\ref{nlVar}).

\item
An initial value of a density fluctuation $\delta_0$ is generated
from the log-normal distribution (\ref{nlPDF}).

\item
The evolution of $\delta$ (at a fixed point) is
calculated using the linear approximations \cite{Peeb1980}
\begin{equation}\label{deltaev}
\ddot{\delta} + 2 \frac{\dot{a}}{a} \dot{\delta} = \frac{4 \pi G}{c^2} \rho \delta.
\end{equation}
\item
Using the Poisson equation, $\nabla^2 \phi$ is calculated and inserted  into (\ref{dDBa})
which is solved from $\chi_i$ to $\chi = \chi_i + \chi(2R)$.

\item
Steps (i)--(v) are repeated so  (\ref{dDBa}) is solved from $\chi = 0$ to $\chi_e$.
\end{enumerate}

A sample of density fluctuations along the line of sight is presented in Figure \ref{fig1}.

\section{Construction of model Len 2}\label{Alen2}

Instead of using directly  the log-normal distribution to 
generate density fluctuations along the line of sight an alternative
approach is used. In \cite{kbMNRAS} it was shown
that compensated structure can also have a log-normal PDF.
Thus, Len 2 generates density fluctuations along the line
of sight using a method discussed in \cite{kbMNRAS}.
The algorithm is as follows:

\begin{enumerate}
\item
First the radius of a void $R_v$ is  generated from the Gaussian distribution 
with the mean of 12 Mpc and the standard deviation of 2 Mpc.
\item Density within the void $\Omega_v$ is generated from the Gaussian distribution with the mean 
 $0.2 \Omega_m$ and $\sigma = 0.27 \Omega_m$.
If a generated in this way $\Omega_v$ is lower than $0.01 \Omega_m$
then the generation is repeated.
If after 6 times it is still less than $0.01 \Omega_m$
then $\Omega_v$ is generated for a uniform distribution between 
 $0$ and $0.01 \Omega_m$. 
If $\Omega_v \geq \Omega_m$
then $\Omega_v$ is generated one more time.
If after 6 times $\Omega_v \geq \Omega_m$ then
its value is chosen for a uniform distribution 
from $0.85 \Omega_m$ to $\Omega_m$. Density contrast is then $\delta = \Omega_v/\Omega_m -1$.
\item Density of the surrounding shell 
$\Omega_s$ is generated from the Gaussian distribution with
the mean of $1.75 \Omega_m$ and $\sigma = 0.7 \Omega_m$.
If $\Omega_s \leq \Omega_m$ then its value
is generated again. If after 6 times 
 $\Omega_s \leq \Omega_m$ then its value is generated for a uniform distribution
between $1.75 \Omega_m$ and $1.95  \Omega_m$.
Density contrast is then $\delta = \Omega_s/\Omega_m -1$.
\item
The condition, that the  structure is compensated implies that the radius of the 
whole structure is

\[R = R_v \left( \frac{\Omega_s - \Omega_v}{\Omega_s - \Omega_m} \right)^{1/3}.\]
\item The angle at which the light ray enters  the structure is generated
from a uniform distribution between $0$ and $0.25\pi$.
\item The evolution of $\delta$ (at a fixed point) was 
calculated using (\ref{deltaev}).
\item 
Integral (\ref{dDBa})
is solved from $\chi_i$ to $\chi_f$ (where $\chi_i$ is the comoving
coordinate of the entry point and $\chi_f$ the point where
the light ray exits the structure).
\item 
Steps (i)--(vii) are repeated so  (\ref{dDBa}) is solved from $\chi = 0$ to $\chi_e$. 
\end{enumerate}

A sample of density fluctuations along the line of sight is presented in Figure \ref{fig1}.
The reason for defining the profile as above
is to have a profile that allows for some kind of randomness (the size and
the density of voids and shells is random), but still because
we use spheres (which suppose to correspond to real wall that surround 
real voids) the fluctuations are not purely random -- the pure randomness
is gone once we decided to use spheres.
Still because of the above construction the 3D
density PDF of this model is almost log-normal.
(see  \cite{kbMNRAS} for a discussion).
If instead of the Gaussian,
the uniform PDF is used (steps (i)--(iii)) then the PDF of $\delta$
is  not log-normal PDF.

\section{Construction of model Len 3}\label{Alen3}

\begin{enumerate}
\item
First the radius of a void $R_v$ is  generated from
a uniform distribution between 0 and 30 Mpc.
\item Density contrast of the void $\delta_v$  is generated
from a uniform distribution between $-0.5$ and $-1$.
\item Integral (\ref{dDBa})
is solved from $\chi_i$ to $\chi_f$ (where $\chi_i$ is the comoving
coordinate of the entry point and $\chi_f$ the point where
the light ray exits the void).
\item
Then the length of a filament $L_f$ is  generated from
a uniform distribution between 0 and 5 Mpc.
\item The density contrast of  the filament $\delta_f$  is generated
from a uniform distribution between $0.5$ and $2.5$.
\item Integral (\ref{dDBa})
is solved from $\chi_i$ to $\chi_f = \chi_i + \chi(L_f)$.
\item 
Steps (i)--(vi) are repeated so  (\ref{dDBa}) is solved from $\chi = 0$ to $\chi_e$. 
\end{enumerate}

A sample of density fluctuations along the line of sight is presented in Figure \ref{fig1}.

\section{Construction of the Swiss Cheese model}\label{Asc}

The Swiss Cheese model considered here is based on the Lema\^itre-Tolman (LT) model \cite{Lema1933,Tolm1934},
which is the simples generalization of the Friedmann models.
The metric of the LT model is

\begin{equation}
{\rm d}s^2 = {\rm d}t^2 - \frac{R,_r^2}{1+2E} {\rm d}r^2 - R^2(t,r) \left({\rm
d}\vartheta^2 + \sin^2 \vartheta {\rm d}\varphi^2 \right),
\end{equation}
where $E(r)$ is an arbitrary function. In the Friedmann limit $E \to -k_0 r^2$
(in a general LT model the `curvature index' is position dependent).
$R(t,r)$ is the areal distance and in the Friedmann limit $R \to r a(t)$ 
(in a general LT model the scale factor is  position dependent).
The evolution (i.e. the generalized Friedmann equation) is

\begin{equation}\label{vel}
\dot{R}^2 = 2E + \frac{2M}{R} + \frac{\Lambda}{3} R^2,
 \end{equation}
$M(r)$ is an arbitrary function defined by the mass density:
 \begin{equation} \label{rho}
 \kappa \rho = \frac {2 M,_r}{R^2 R,_r}.
 \end{equation}
From (\ref{vel}) the age of the Universe, $\tau$, is 
 \begin{equation}\label{tbg}
\tau(r) = t- t_B(r) = \int\limits_0^R\frac{{\rm d} \widetilde{R}}{\sqrt{2E +
2M/\widetilde{R} +
 \frac{1}{3}\Lambda \widetilde{R}^2}},
 \end{equation}
where $t_B(r)$ is an arbitrary function and in the Friedmann model $t_B \to$ const
(in a general LT model the age of the universe is position dependent).
There are three arbitrary functions in the LT model $t_B(r), M(r)$, and $E(r)$,
however only 2 are physically independent.
For a review of different applications of the LT models see
\cite{Kras1997,PlKr2006,BKHC2009}.
Here we will assume that the model is defined by the assumption that the age of the universe
is everywhere the same, $t_B = 0$ and the function $M(r)$ is

\begin{equation}\label{Mm1}
M = M_0 +  \left\{ \begin{array}{ll}
M_1 \ell^3 & {\rm ~for~} \ell \leqslant x_a, \\
M_2 \exp \left[ -  \left( \frac{ \ell - 2x_a}{x_a} \right)^2 \right] & {\rm ~for~} x_a \leqslant  \ell \leqslant 3x_a \\ 
-M_1 (\ell - 4x_a)^3 & {\rm ~for~} 3x_a \leqslant  \ell \leqslant  4x_a,  \\
0 &  {\rm ~for~} \ell \geqslant 4x_a, 
\end{array} \right.
\end{equation}
where $\ell = r$/kpc, $M_0 = (4 \pi G /3c^2) \rho_{b} \ell^3$,
$\rho_{b} =  \Omega_m \frac{3H_0^2}{8 \pi G}$,
$\Omega_m = 0.262$, $H_0=71.4$ km s$^{-1}$ Mpc$^{-1}$,
$M_1 = x_a^{-3} M_2 {\rm e}^{-1.5}$, 
$M_2= 3.29 \times 10^{10}$ kpc and $-2.91 \times 10^{11}$
for regions A and B respectively, 
$x_a= 5\times 10^3$ and $15\times 10^3$, for regions A and B respectively.
Region A and B are matches alternately at $r = 4 x_a$.
 The above profile was chosen for the following reasons:
it behaves like a FLRW model for $\ell \leqslant x_a$ but with lower
density than outside, then for $x_a \leqslant  \ell \leqslant 3x_a$
we have a transition region, and a cubic behavior for $ 3x_a \leqslant  \ell \leqslant  4x_a$  allows for a smooth matching to the background values. 
Density fluctuations along the line of sight are presented in Figure \ref{fig1}.
It should now be clear why we have selected the above set of functions.
These functions allow us to obtain the present-day density profile as presented in Figure \ref{fig1}.
This model has thus 2 advantages: 1) it allows to calculate the evolution and the
distance exactly, without additional approximations, 2) the mean of the density fluctuations along the line
of sight is zero, i.e. as in Len 1 model and unlike in Len 2 and Len 3 models.

When constructing a Swiss Cheese model, the junction
conditions need to be satisfied. Here we match the LT inhomogeneity (holes)
to the Friedmann background (cheese).
The LT patches are placed so that
their boundaries touch wherever a light ray exits one inhomogeneous patch.
Thus, the ray immediately enters another LT patch and does not 
spend any time in the Friedmann background. 
To match the LT patch to the Friedmann background across a comoving
spherical surface, $r =$ constant, the conditions are: 1) the mass inside the
junction surface in the LT patch is equal to the mass that would be inside
that surface in the homogeneous background;
2) the spatial curvature  at the junction surface is the same in both
the LT and Friedmann models, which implies that $E(r) = - k_0 r^2$
and $E' = - 2 k_0 r$; 3) the bang time and  $\Lambda$ must be continuous across the junction. As seen, by the construction these conditions are satisfied.

Finally, the redshift is given by  \cite{BKHC2009}
	
\begin{eqnarray}
&& \frac{{\rm d} r}{{\rm d} z} = \pm \frac{1}{1+z} \frac{\sqrt{1+2E}}{
\dot{R}'}, \nonumber \\
&& \frac{{\rm d} t}{{\rm d} z} = \frac{1}{1+z} \frac{R'}{
\dot{R}' }.
\label{redrel}
\end{eqnarray}
Only radial geodesic (i.e. the geodesics that propagate through the origin) will be considered.
In such a case the shear of light bundle is zero \cite{BrTT2007,BrTT2008}.
Thus, the junction of null geodesics only requires the matching of one
of the components of $k^\alpha$ as the other one is given from $k^\alpha k_\alpha = 0$.

\end{document}